\documentclass[a4paper]{jpconf}
\usepackage{graphicx}
\begin{document}
\title{Background rejection method for tens of TeV gamma-ray astronomy applicable to wide angle timing arrays}

\author{A Sh M Elshoukrofy$^{1,2}$, E B Postnikov$^{1,*}$ and L G Sveshnikova$^{1,**}$}
\address{$^1$ Lomonosov Moscow State University Skobeltsyn Institute of Nuclear Physics (MSU SINP), Leninskie gory 1(2), GSP-1, Moscow, 119991, Russia}
\address{$^2$ Faculty of Science, Damanhour University, El-Gomhouria St. 22516, Damanhour, El Beheria, Egypt}

\ead{$^*$evgeny.post@gmail.com, $^{**}$tfl10@mail.ru, abeershehatamahmoud@yahoo.com}

\begin{abstract}
A 'knee-like' approximation of Cherenkov light Lateral Distribution Functions, which we developed earlier, now is used for the actual tasks of background rejection methods for high energy (tens and hundreds of TeV) gamma-ray astronomy. In this work we implement this technique to the HiSCORE wide angle timing array consisting of Cherenkov light detectors with spacing of 100 m covering 0.2 km$^2$ presently and up to 5 km$^2$ in future. However, it can be applied to other similar arrays. We also show that the application of a multivariable approach (where 3 parameters of the knee-like approximation are used) allows us to reach a high level of background rejection, but it strongly depends on the number of hit detectors.
\end{abstract}

\section{Introduction}
\label{Introduction}
The ground-based high energy gamma-ray astronomy became the most dynamically developing field in very high energy astrophysics over the last decade. Detection of gamma rays that come from galactic and extragalactic sources has some advantages compared to detection of charged cosmic rays. First, gamma rays keep their emission direction from the source, and second, they are easy to detect unlike the other type of particle with the same property, neutrinos. Most of the knowledge about sources of high energy gamma rays has been obtained from arrays of Imaging Atmospheric Cherenkov Telescopes (IACT) \cite{1}. The method proposed by A M Hillas in 1985 \cite{2} allows distinguishing of the extensive air showers produced by gamma rays from those initiated by charged cosmic rays. However, to access the energy range from 10 TeV up to several 100 TeV a detection area of about 10 km$^2$ or greater is required. 

Alternatively to imaging gamma-ray telescopes, the shower-front sampling technique allows installing setups of large effective area and also naturally provides large viewing angles of the instrument. Its operating principle is based on sampling the density and arrival time of the air shower front with distributed arrays of detector stations \cite{3, 4, 5}. However, a very difficult problem is the discrimination of gamma ray initiated showers against the background of cosmic ray showers \cite{6}. To overcome this problem, in the experiment TAIGA ``Tunka Advanced Instrument for cosmic ray physics and Gamma-ray Astronomy'' a wide-angle Cherenkov array Tunka-HiSCORE was proposed to be combined with narrow-angle imaging detectors Tunka-IACT \cite{7}. 

HiSCORE is an array of wide-angle Cherenkov stations, currently composed of 28 detector stations distributed in a regular grid over a surface area of 0.25 km$^2$ with an inter-station spacing of about 100 m. The detector stations measure the light amplitude and arrival time delay over a distance of a few hundred meters. Signal amplitude and time delay in every triggered station are used for the reconstruction of shower direction, core position, and primary energy. Discrimination of gamma-induced showers against the background of charged cosmic rays in the HiSCORE experiment is a 3-stage process. At the first step a background flux is suppressed due to a very good arrival direction resolution $d\Theta$=0.1--0.4$^\circ$, because a  background flux is proportional to the intensity of cosmic rays and solid angle of observation, $d\Omega$$\sim$$(d\Theta)$$^2$. At the second step, which is the aim of the present paper, an event by event selection based on the parametric analysis of LDF (lateral distribution  functions) of events is planned. After the commissioning of the IACT as a part of the array TAIGA-HiSCORE, the hybrid timing/imaging method of selection will be used. Preliminary results of simulations \cite{8} show that the joint operation of the HiSCORE array and a net of IACTs can be very effective. 

The idea of using the difference in LDF for identification of particles with different depth of shower maximum appeared decades ago. It was noted and explained that the depth of shower maximum is correlated with the steepness of the LDF \cite{9}. The steepness was reconstructed from the ratio of the light intensities at different distances (for example, at 50~m / 150~m, or 30~m / 125~m depending on features of experimental  arrays \cite{3}). In the TUNKA experiment a more complicated steepness parameter $bxy$ was proposed \cite{10}.    

The aim of this paper is to study a possibility of a background rejection based on difference in LDF of Cherenkov light for gamma induced and hadron induced showers. For that purpose we apply a 'knee-like' approximation of LDF, which we developed earlier \cite{11}. This approximation is appropriate for effective parametric analysis of LDF as well as for application of multivariable approach in conditions of the HiSCORE experiment.

\section{'Knee-like' approximation and parametric analysis for the true LDF}
\label{sec-2}
For parameterization of the LDF at a distance $R$, $Q(R)$, we proposed \cite{11} a simple function designated as a ‘knee-like approximation’. It's a one-variable function depending also on five parameters $C$, $\gamma_1$ , $\gamma_2$ , $R_0$, and $\alpha$:
\begin{equation}
\label{function-1}
Q(R) = CR^{\gamma _1}(1+(\frac{R}{R_0})^\alpha)^\frac{\gamma_2}{\alpha}
\end{equation}
In spite of the significant diversity, for all the events the LDF has a specific knee-like structure. A flat direct Cherenkov light disk ($Q(R){\sim}R^{\gamma_1}$ with the slope  $\gamma_1$$\sim$--0.7$\div$0.2) is extended up to the distance $R_0$ (a 'knee' position). In the region beyond $R_0$, $Q(R){\sim}R^{\gamma_1+\gamma_2}$ and a photon density decreases very steeply with the slope $\gamma_1$+$\gamma_2$$\sim$--2. $R_0$ ranges from 50 m to 200 m and depends strongly on the distance from the detector to the shower maximum. The parameter $\alpha$ is responsible for the sharpness of the knee providing a smooth transition between the power laws at the knee point. This parameter is the main distinction of our approximation because it is appropriate for describing the both cases: a very smooth transition from the central part to the periphery across the knee ($\alpha$$<$3), and a very sharp transition (usually $\alpha$$>$50). The value of $\alpha$ ranges from 0 to infinity, its logarithm can classify the LDFs as belonging to 2 separate clusters: `smooth' ones and `sharp' ones.

\begin{figure}[h]
\begin{minipage}{20pc}
\includegraphics[width=20pc]{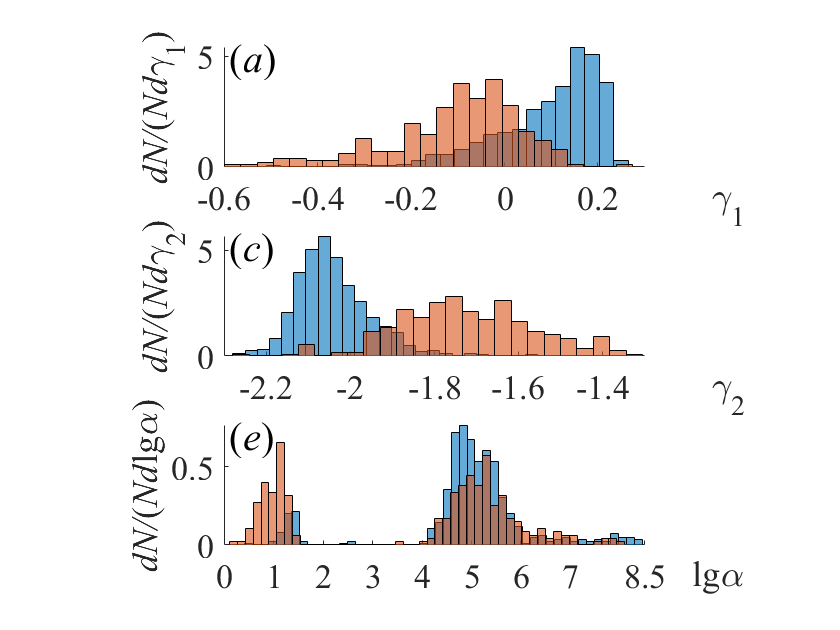}
\end{minipage}
\begin{minipage}{20pc}
\includegraphics[width=20pc]{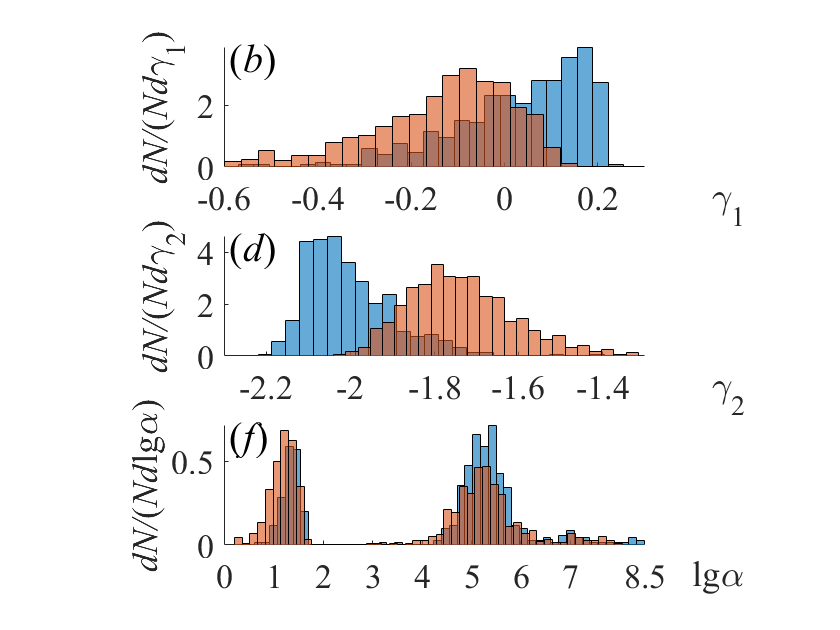}
\end{minipage} 
\caption{\label{figure1}Comparison of distributions of parameters $\gamma_1$ (\textit{a})--(\textit{b}), $\gamma_2$ (\textit{c})--(\textit{d}), lg$\alpha$ (\textit{e})--(\textit{f}) for gamma rays (blue) and protons (orange) at energy 30 TeV (left panel) and 100 TeV (right panel), angles 25-50$^\circ$.}
\end{figure}

Parameters of approximation $\gamma_1$, $\gamma_2$, $R_0$, and $lg\alpha$ depend on the energy and type of primary particle and allow separating proton and gamma induced showers for the true LDF in the wide energy interval 30--3000 TeV and zenith angle range 0--50$^\circ$. The true LDF is a designation of functions simulated with a space step of 5 m. In figure \ref{figure1} we present distributions of parameters $\gamma_1$, $\gamma_2$, $lg\alpha$ for gamma ray and proton induced showers with energy 30 TeV and 100 TeV, in the interval of zenith angles 25-50$^\circ$. The brown area marks common parts of gamma ray and proton distributions. We can see for all of these parameters that gamma ray and proton distributions are visually separated from each other, and this fact can be used to discriminate between sorts of particles. A discrimination between gamma rays and protons in a 3-dimensional space of these parameters using quadric separating surfaces in quadratic discriminant analysis (QDA \cite{12}) leads to large value of efficiency of proton rejection ($\sim$100 times).
\section{Rejection methods for real condition}
\label{sec-3}
The real conditions of registration suppose that we select events with a small number of triggered stations (4--5) to decrease the energy threshold. In that case we also do not know the core position and therefore insert in a fitting function two additional unknown parameters ($X$-core, $Y$-core). The knee-like approximation of LDF of an array of our real 28 detectors strongly depends on the number of hit detectors, since only few detectors have a signal above noise level. 

\begin{figure}[h]
\includegraphics[width=28pc]{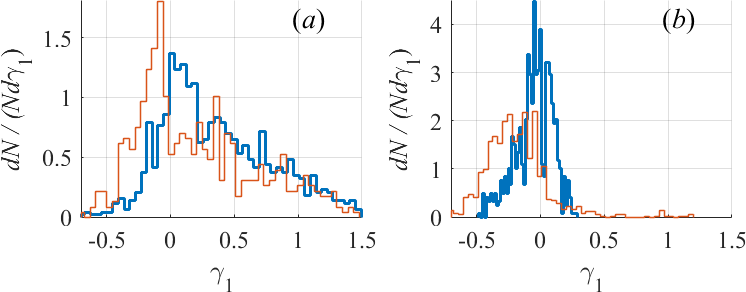}
\includegraphics[width=28pc]{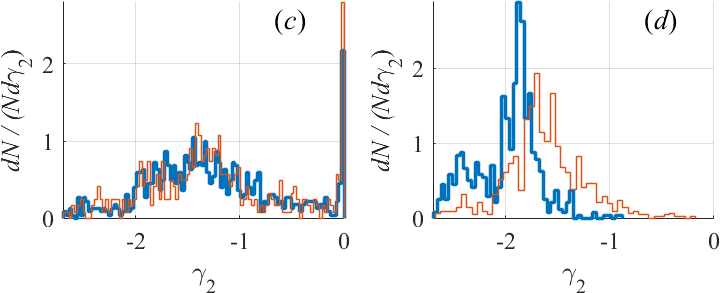}
\includegraphics[width=28pc]{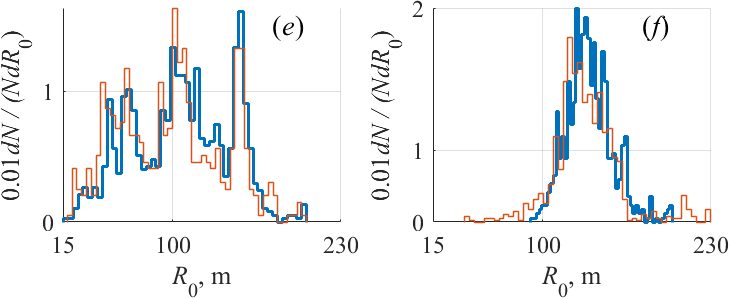}
\includegraphics[width=28pc]{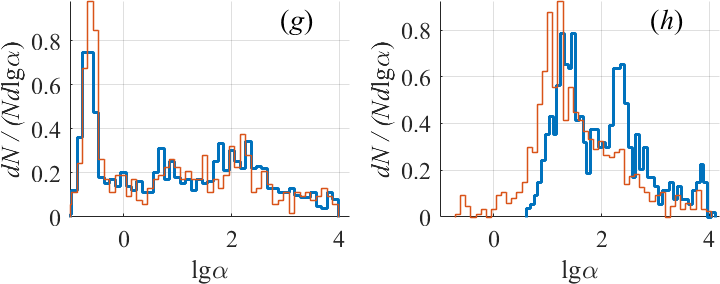}
\caption{\label{figure2}Comparison of distributions of parameters $\gamma_1$ (\textit{a})--(\textit{b}), $\gamma_2$ (\textit{c})--(\textit{d}), $R_0$ (\textit{e})--(\textit{f}), and lg$\alpha$ (\textit{g})--(\textit{h}) for gamma rays (blue line) and background (proton and helium mixture, orange line) for a different number of hit detectors (left panel -- 7 hit detectors; right panel -- 13--28 hit detectors). Energy 30--3000 TeV, zenith angles 29--42$^\circ$.}
\end{figure}

\begin{figure}[h]
\begin{minipage}{18pc}
\includegraphics[width=18pc]{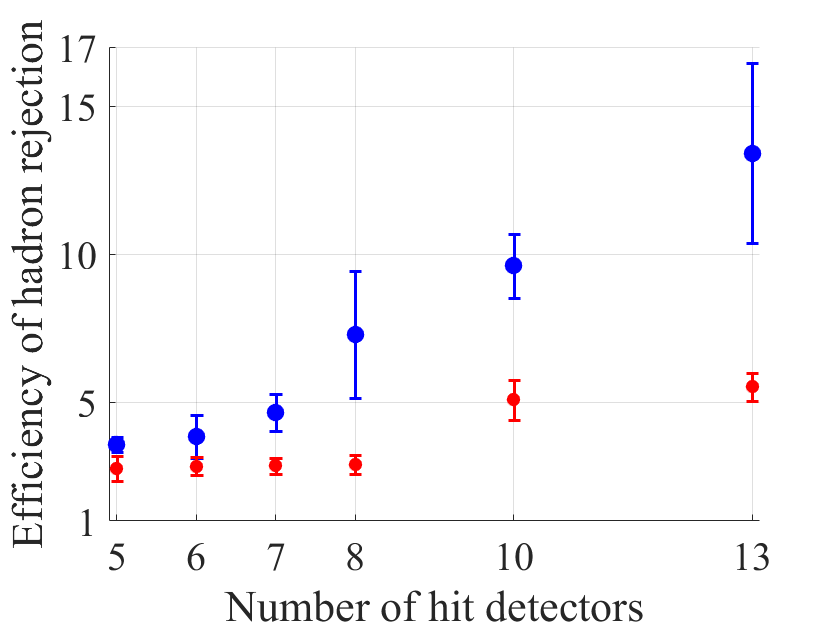}
\end{minipage}
\begin{minipage}{20pc}
\caption{\label{figure3}Efficiency of rejection of hadron background (background before rejection / background after rejection) vs number of hit detectors. Only the difference in LDFs for gamma and hadron induced showers is used. Red points -- single parameter rejection, blue points -- 3-parameter rejection by quadratic discriminant analysis.}
\end{minipage} 
\end{figure}

Our previous study \cite{13} showed that different steps of reconstruction (core position reconstruction and particle identification) require different versions of the knee-like LDF fit with different number of fitting parameters. For the core position reconstruction in \cite{13} we reduced a number of fitting parameters from 5 to only 2 and have shown that the reduced version of fit has a distinctive advantage when we work with a small number of triggered detectors. However, it reduces the diversity of the LDF to a fixed set of curves. Therefore, first we apply the 2-parameter fit as in \cite{13}, and then investigate the quality of gamma/background discrimination using the 5-parameter fit. In figure \ref{figure2} we present the comparison of distributions of parameters $\gamma_1$, $\gamma_2$, $R_0$, and $lg\alpha$ for gamma rays and background (proton and helium mixture) for a different number of hit detectors. Events were simulated with the power low spectrum with the slope $-1.6$ in the energy interval 30--3000 TeV for zenith angles 29--42$^\circ$. These conditions were designed for the expected sample of gamma rays from the Crab Nebula and corresponding proton background. In figure \ref{figure3} the efficiency of rejection of hadron background is given depending on the number of hit detectors.

Figures \ref{figure2}, \ref{figure3} demonstrate a good rejection of background (proton and helium mixture), though it depends on the number of hit detectors. For a small number of hit detectors ($\sim$7--8) the difference of distributions is slight (weak rejection), whereas for a large number of detectors ($>\sim$8) a noticeable difference of distributions appears. If we apply the multivariable approach (QDA, section \ref{sec-2}), the background rejection efficiency (background  before rejection / background after rejection) varies from 4 to 14 depending on the number of hit stations (figure \ref{figure3}). It is worth to stress that the rejection based on the LDF difference between gamma rays and hadrons is only a part of the complete rejection procedure to be implemented in TAIGA-HiSCORE (section \ref{Introduction}), therefore, it will lead to a greater improvement of the integral sensitivity. 

\section{Conclusions}
\label{sec-6}
This study demonstrated that a 'knee-like' approximation of Cherenkov LDF, which we developed earlier, can be used in the actual tasks of background rejection for high energy gamma-ray astronomy. The approximation gives the possibility to perform a multivariable parametric analysis of LDF and allows separating proton and gamma induced showers. For real condition of registration in the HiSCORE experiment a good discrimination of gamma rays against proton/helium background can also be made, but only for a large number of triggered stations.
\ack
We thank E E Korosteleva for the simulated data bank that was used in our analysis. The study was supported by the Russian Foundation for Basic Research, project no.  16-29-13035.

\end{document}